\documentclass[11pt,preprint,flushrt]{aastex}

\def\eqq#1{Equation~(\ref{#1})}
\newcommand\etal{{\it et al.\/}}

\newcommand\Pp{P^{\prime}}
\newcommand\Ppp{P^{\prime \prime}}
\newcommand\Pppp{P^{\prime \prime \prime}}
\newcommand\Deltap{\Delta^{\prime}}

\begin{document}

\slugcomment{\$Revision: 1.18 $ $ \$Date: 2010/01/13 15:16:41 $ $ submitted to PASP}

\title{Noise and bias in square-root compression schemes} 
\author{Gary M. Bernstein\altaffilmark{1},
Chris Bebek\altaffilmark{2},
Jason Rhodes\altaffilmark{3,4},
Chris Stoughton\altaffilmark{5},
R. Ali Vanderveld\altaffilmark{3,4},
\& Penshu Yeh\altaffilmark{6} }
\altaffiltext{1}{Dept. of Physics and Astronomy, University of Pennsylvania,
Philadelphia, PA 19104}
\email{garyb@physics.upenn.edu}
\altaffiltext{2}{Lawrence Berkeley National Laboratory, 1 Cyclotron Road, Berkeley, CA, 94720}
\altaffiltext{3}{Jet Propulsion Laboratory, California Institute of Technology, Pasadena, CA 91109, U.S.A.}
\altaffiltext{4}{California Institute of Technology, 1200 East California Boulevard, Pasadena, CA 91125, U.S.A}
\altaffiltext{5}{Fermi National Accelerator Laboratory, PO Box 500, Batavia, IL 60150}
\altaffiltext{6}{Goddard Space Flight Center, Greenbelt, MD 20771}

\begin{abstract}
We investigate data compression schemes for proposed all-sky diffraction-limited visible/NIR sky surveys aimed at the dark energy problem.  We show that lossy square-root compression to 1 bit of noise per pixel, followed by standard lossless compression algorithms, reduces the images to 2.5--4 bits per pixel, depending primarily upon the level of cosmic-ray contamination of the images.  Compression to this level adds noise equivalent to $\le 10\%$ penalty in observing time.  We derive an analytic correction to flux biases inherent to the square-root compression scheme.  Numerical tests on simple galaxy models confirm that galaxy fluxes and shapes are measured with systematic biases $\lesssim10^{-4}$ induced by the compression scheme, well below the requirements of supernova and weak gravitational lensing dark-energy experiments.  An accompanying paper \citep{Ali} bounds the shape biases using realistic simulated images of the high-Galactic-latitude sky.  The square-root preprocessing step has advantages over simple (linear) decimation when there are many bright objects or cosmic rays in the field, or when the background level will vary.
\end{abstract}


\section{Introduction}
The unexpected discovery of the acceleration of the Hubble expansion of the Universe has inspired numerous proposals for large-scale observational projects to seek further constraints on this ``dark energy'' problem \citep[see][for a review of dark energy]{FTH}.  Several such proposals involve space-based visible or NIR imaging surveys of a substantial fraction of the full celestial sphere at resolution near the diffraction limit of 1--2~meter telescopes.  Such surveys would also be extraordinarily valuable for general astrophysics, but the amount of data involved is large.  The number of pixels in a survey of a fraction $f_{\rm sky}$ of the sky, with pixel scale $p$ and $N_{\rm exp}$ exposures per sky location, will be
\begin{equation}
N_{\rm pix} = 2.7\times 10^{14} \left({ p \over 0\farcs1 }\right)^{-2} { f_{\rm sky} N_{\rm exp} \over 5}.
\end{equation}
A crate filled with disk drives is all that is needed these days for transmission of hundreds of terabytes of data between terrestrial sites.  The transmission of this data from a spacecraft at the Earth-Sun L2 Lagrange point 1.5~million km away is, however, a substantial expense.  If the data require an average of bpp bits per pixel to transmit, with a telemetry rate $r$, and a survey duration of $T$, then the number of hours of downlink per day required will be
\begin{equation}
\mbox{downlink time} ={\rm bpp}\times 0.33 {\rm hrs \over day}
{ f_{\rm sky} N_{\rm exp} /T \over 1.25/{\rm yr}}
\left({r \over 150\,{\rm Mbps}}\right)^{-1} \left({ p \over 0\farcs1 }\right)^{-2}.
\end{equation}
To avoid excessive resource use both on the spacecraft and ground stations,
the data volume must be reduced well below the
16 bits per pixel that is typical for digitization of astronomical array detectors.
In this paper we investigate means for compressing astronomical imaging data well below this, ideally ${\rm bpp} \lesssim 4$.

Significant compression of the detector output signals will require a lossy compression step to reduce the bit depth of the noise---sometimes called ``pre-processing''---followed by a lossless algorithm to reduce the mean bit depth to the signal entropy level.  Because most of the pixels in an extragalactic visible/NIR exposure will be close to the sky level, it is clear that the primary determinant of the final bpp value will be the number of noise bits retained by the lossy step in pixels near sky level---extensive demonstration of this for ground-based images is done by \citet{PSW}.
Optical astronomers are averse to lossy compression for fear of losing the information from their dearly acquired photons or fear of inducing biases in high-precision measurements.  It should be pointed out, however, that lossy compression is already being performed on essentially all data when the analog detector outputs are digitized.  Analog-to-digital gains are usually set, however, so that the errors induced by digitization are below the read noise of the detector system.  In radio astronomy it is well known that digitization to only 1 or 2 bits per sample can capture nearly all the signal information when $S/N$ per sample is low \citep{Weinreb,TMS}.  The critical parameter for information loss is the number of bits of noise retained, {\it i.e.} the ratio $b=\sigma_N/\Delta$ of the signal noise to the digitization step size. We seek a lossy compression scheme which holds $b$ fixed over the full dynamic range of the detector.  When the signal noise is dominated by shot noise from the detected photoelectrons, it is straightforward to show that a {\em square-root compression scheme} holds $b$ constant.  The application of square-root compression to Poisson-dominated data has been discussed by \citet{GowenSmith} and is part of the signal chain for the {\it James Webb Space Telescope} \citep{JWST} and {\it Supernova Acceleration Probe} \citep{LinMarriner} spacecraft designs.  Further discussion is in \citet{SWP}. In \S\ref{algo} we review the algorithms of such a scheme and \S\ref{noisesec} shows that the square-root compression algorithm increases the noise by a fraction $1/24b^2$ for cases of interest.

Once a lossy compression algorithm is determined to reduce the data volume with acceptably small degradation of the noise level, it is critical to determine whether the codec process will induce any bias on the astronomical measurements from the data.  Future dark energy experiments will require very demanding, high-precision measurements.  For use of Type Ia supernovae as distance indicators, systematic errors in flux measurements must be $\ll 1\%$.  For determination of galaxy redshifts using the ``photo-z'' method, similarly small systematic errors must be present in galaxy photometry.  We target codec-induced systematic errors in flux to be $<10^{-3}$ so that they contribute negligibly to the photometric error budget.

Even more demanding are programs for weak gravitational lensing (WL) measurements, which depend upon determination of the shapes (ellipticities) of $\gtrsim 10^9$ galaxies on the sky.  \citet{HTBJ} and \citet{AmaraRefregier} both conclude that measurements of galaxy ellipticities must have systematic errors of $\le 10^{-3}$ in order to remain subdominant to statistical errors in the WL experiments.  
Because there are many possible sources of shape measurement error---primarily uncorrected instrumental image distortions---we would like to have shape biases induced by the codec be a minor contributor to the shape error budget, {\it i.e.} the shift $\delta e_i$ in each component $e_i$ of galaxy ellipticity should on average be
$<10^{-4} e_i$.  The goal of this paper is to examine the biases induced by square-root compression down to $b=1$ bit of noise per pixel, and to show how these biases can be reduced to negligible levels.  A companion paper \citep{Ali} tests for codec-induced shape bias on realistic simulations of space-based CCD imaging.

Many lossy image compression algorithms are based on decomposition of 2d images over a function library, followed by truncation of the terms in the library---for example wavelet transforms \citep{White,Press} or the JPEG algorithms.  Even more radical compression is possible with ``compressed sensing'' techniques if the data are known to be sparse in some library \citep{Candes}.  For WL applications we are wary of lossy compression algorithms that examine groups of pixels and might hence induce correlated codec errors over multiple pixels.  We therefore focus here on a lossy step that compresses data strictly on a pixel-by-pixel basis.  If single-pixel codec errors are unbiased and completely uncorrelated between pixels,  object shape measurements should be unbiased.  We investigate the single-pixel codec in this work,  leaving open the question of whether grouped-pixel algorithms can meet the shape bias requirements of weak lensing surveys while obtaining better compression than single-pixel codecs. 

In \S\ref{biassec} we show that there is a significant bias in the mean values reconstructed by the square-root class of codecs.  We also show, however, that a straightforward adjustment of the decoded values can remove this bias to remarkable accuracy, largely independent of the details of the probability distribution of the raw data.  We therefore expect the codec to produce unbiased flux and shape measurements, which we test with simplified galaxy images in \S\ref{testsec}.  \citet{Ali} test for shape-measurement bias on data that is a realistic simulation of images expected from the {\it Joint Dark Energy Mission (JDEM)}\footnote{\tt http://jdem.gsfc.nasa.gov} CCD imager.

Finally in \S\ref{bppsec} we compress realistic images with the lossy square-root algorithm followed by standard lossless algorithms, and measure the number of bits per pixel required to transmit the fully compressed images.  In the conclusion we summarize our findings, namely that it is possible to compress {\it JDEM}-like images to 3--4 bits per pixel with only 4\% penalty in image noise, and to reconstruct them such that biases in flux and shape measurements induced by the codec are not detected, down to $\approx 10^{-4}$ levels.

\section{Compression algorithm}
\label{algo}
We consider most generally a compression algorithm that maps the digitized detector output value $N$ into a compressed code $i$ whenever $N_i-\Delta_i/2 < N \le N_i+\Delta_i/2$.  
Here $N_i$ and $\Delta_i$ are the center and width, respectively, of the data range that codes to compressed value $i$.  Note that $N_{i+1}-N_i=(\Delta_i+\Delta_{i+1})/2.$ We will assume that $N$ is a debiased value that is linearly related to the estimated number of photoelectrons $e$ by $N=e/g$, where $g$ is the (inverse) gain of the system.
We will assume initially that decompression consists of mapping $i\rightarrow N_i$, {\it i.e.} we restore code $i$ to the mean of the input values that it codes.  In a later section we will suggest an alteration to the decompressed values which reduces the bias in reconstructed values.  

Any vector of increasing $N_i$ (or equivalently a vector of $\Delta_i>0$) defines a codec algorithm.  We expect the bias and noise characteristics of the codec to depend primarily on the the number of bits of noise retained in the compression, which we define as
\begin{equation}
b = \sigma_N / \Delta,
\end{equation}
where $\sigma_N$ is the RMS noise level of a pixel and $\Delta$ is the typical $\Delta_i$ for the codes $i$ that span the noise distribution of the pixel.  Many astronomical detectors have noise described by
\begin{equation}
\sigma_N^2 = [e + (RN)^2]/g^2 = (N+c)/g, \qquad c\equiv (RN)^2/g,
\end{equation}
with some read noise level $RN$ added to the Poisson variance of the photoelectrons.  In this case, if we wish to keep $b$ independent of the signal level, we require
\begin{eqnarray}
\Delta = \left[{di \over dN}\right]^{-1} & = & \sigma_N/b  \\
\Rightarrow \quad
{di \over dN} & = & {b \over \sqrt{(N+c)/g}} \\
\Rightarrow \quad
i & = & \sqrt{k(N+c)}, \qquad k\equiv 4gb^2.
\label{delta1}
\end{eqnarray}
Since the compressed code $i$ must be integral, we can more precisely express the nominal square-root compression algorithm as
\begin{equation}
i = \left[ \sqrt{k(N+c)}\right] = \left[2b\sqrt{e+RN^2}\right],
\label{encode}
\end{equation}
where, in this equation only, $[x]$ is the nearest integer to $x$.  \citet{GowenSmith} discuss more general offsets under the square root, but we stick to these values since they hold $\sigma_N/\Delta$ fixed across the dynamic range.

\subsection{Range step sequences}
\label{stepseq}
If we define
\begin{equation}
\Deltap_i = \Delta_i - \Delta_{i-1}
\end{equation}
then from \eqq{delta1} we find that the square-root codec has
\begin{eqnarray}
\Delta_i & \approx & {2\sqrt{k(N+c)} \over k} = {2 i \over k} \\
\Deltap_i & \approx & {2 \over k} = {1 \over 2b^2g},
\end{eqnarray}
{\it i.e.} the width $\Delta_i$ of each code range increases at a constant rate.  If the raw data $N$ are integral, then these $\Delta_i$ and $\Deltap_i$ must also be integral.  If for example we desire $b=1$ bit of noise and have gain $g=0.5$ $e$/ADU, then $\Deltap_i=1$ produces the desired compression algorithm.  
When $1/2b^2g$ is non-integral, however, we will more generally obtain $\langle \Deltap_i \rangle=1/2b^2g$.  In fact we can define a class of compression algorithms as follows:
\begin{enumerate}
\item Choose a sequence of integers $\{s_1, s_2, \ldots, s_n\}$ that will repeat cyclically to give the incremental range steps $\{\Deltap_1, \Deltap_2, \ldots\}$.
\item Define the parameter $b$ via 
\begin{equation}
b^2 = {1 \over 2g \langle s_i \rangle}.
\end{equation}
\item Choose a range $\Delta_0$ for the first compression code that extends from $N=0$ to
$N=(RN)/gb$.  [The codec algorithm can be extended to small negative values of $N$ if desired using the fixed range width $\Delta_0$ for small negative $i$ values.]
\end{enumerate}
Then the codec defined by this range-step sequence will have nearly constant number $b$ of bits per pixel noise level under the Poisson$+$read noise data model.  The more nearly uniform the values $s_i$ are, the better will be the codec performance, as shown below.
For example we can produce a codec with $b=1$ with $g=1$ by using code steps $\Deltap_i=\{0,1,0,1,0,1,\ldots\}$.   We will refer to this as a ``$g=1$, $\{0,1\}$ codec.''

\subsection{Data model}
When considering the compression to act upon integer data, we denote the probability of the raw pixel value being $N$ as $P_N$.  For notational convenience we will assume that there exists a continuous and differentiable function $P(x)$ over $x\ge0$ which matches $P_N$ at $x=N$.  For example the Poisson formula for $P_N$ is easily extended to non-integral values with a gamma function.

In the CCD data the Poisson signal has a Gaussian read noise added to it.  We will assume the read noise has zero mean. 
Before digitization we can hence consider the pixel photoelectron count $e$ to be a real number, with probability distribution $P(e)$.  In the limit of high gain $g$ we can consider the detector output to be a real number $e$ or $N$; at finite gain, $P_N$ is again defined only at integral $N$, but we can assume there exists some analytic extension to the full real domain $P(N)$.

Most of our results are general to any well-behaved $P(N)$ or $P(e)$, and we will note when the assumption of Poisson or Poisson$+$Gaussian noise models are important.

\section{Codec noise}
\label{noisesec}
A raw pixel value $N$ will incur an error $\delta N = N_i - N$ upon being encoded and decoded through code $i$.  If the input values are uniformly distributed across the range of code $i$, then the RMS error induced by the codec is $\sigma_{\rm codec} = \Delta_i / \sqrt{12}$.  If we make the further simplifying assumptions that the coding error is uncorrelated with the raw $N$ and that $\Delta$ is nearly constant for all $i$ accessed by $P(N)$, then the variance of the output is
\begin{equation}
\sigma^2_{\rm out} = \sigma_N^2 + \sigma^2_{\rm codec} = \sigma^2_N \left(1+ {1 \over 12b^2}\right).
\label{noiseeq}
\end{equation}
The integration time required for a shot-noise-limited astronomical exposure to reach fixed signal-to-noise ratio ($S/N$) will be inflated by 
\begin{equation}
{\sigma^2_{\rm out} \over \sigma_N^2} = 1 + {1 \over 12b^2}.
\end{equation}
If we want the codec to cause $<10\%$ penalty in exposure time, we hence need $b\ge0.91$.  In this paper we will focus our attention on this question: {\em does a square-root algorithm with $b\ge1$ cause any bias or degradation in astronomical measurements beyond the expected $<4\%$ increase in noise level?}  

A floor on the bandwidth required to transmit real images is given by the bpp found for pure-noise images.  The minimum bpp required to for lossless compression of a sequence of codes $i$ with probabilities $p_i$ is the Shannon entropy $H=-\sum p_i \log_2 p_i$.
For nearly-Gaussian probability distributions digitized to $b$ bits per $\sigma$, the Shannon entropy is $H\approx \log_2 \sqrt{2\pi e} + \log_2 b$\citep[e.g.][]{Romeo}.  For $b=1$ a numerical evaluation of the Shannon entropy shows that $\ge2.1$ bpp will be required.
We construct a 16-bit integer FITS image consisting of random noise with $\bar e=40$ and $RN=4$, apply square-root compression with $b=1$, then apply lossless {\tt bzip2} or CCSDS 121B \citep{ccsds} compression.  The resulting image requires 2.4--2.5 bits per pixel, within 0.3--0.4 bpp of the Shannon limit.  We can therefore consider 2.4 bits per pixel to be a floor on the telemetry required when using a pixel-by-pixel lossy compression step, subject to $\le10\%$ integration-time penalty from codec errors. We note that the ``raw'' data digitized at $g=1$ would have $b=7.5$ and require 2.9 more bits per pixel.  
We verify that with square-root compression enabled, the bpp for pure-noise images is largely independent of $\bar e$,  $RN$, and $g$,  as expected.  

\section{Reconstruction bias}
\label{biassec}
The expectation value of the raw data $\bar N=\langle N \rangle$ should be unchanged after compression and decompression.  We will denote by $\delta \bar N $ the bias in the expectation value of the data after the codec operates.  In practice this bias is not a problem if it is independent of $\bar N$, since it is easily corrected.  If $\delta \bar N $ varies with $\bar N$ and/or with the details of the $P(N)$ distribution, then the response of the detector effectively becomes nonlinear and/or ambiguous to correct, which can bias photometry and shape measurements.

Any codec bias could of course be corrected if the raw distribution $P_N$ is known at each pixel.  But the mean level $\bar N$ will vary from pixel to pixel due to background gradients and sources, so there is no single $P_N$ that applies to the entire image.  We would prefer to debias the data by decompressing code $i$ to $N_i-\delta_i$, where $\delta_i$ is some small correction that is independent of the raw probability distribution.  There is no guarantee that such a solution exists but we will show that a good approximation does exist for cases when $P(N)$ spans multiple codes.

\subsection{Real-valued data}
We begin with the case in which the raw value $N$ can be considered a continuous real variable.  This case is instructive if not necessarily realistic.  The center and width $N_i$ and $\Delta_i$ of the range for each code $i$ can be set to any positive real numbers.
The codec bias can be expressed as
\begin{equation}
\delta \bar N = -\sum_i \int_{-\Delta_i/2}^{\Delta_i/2} du\,
u\, P(N_i+u),
\label{bias1}
\end{equation}
where $u=N-N_i$ is the distance from the center of code step $i$.  
If we can find a set of $\delta_i$ such that
\begin{equation}
\sum_i \int_{-\Delta_i/2}^{\Delta_i/2} du\, \delta_i P(N_i+u) = \delta \bar N
\label{bias2}
\end{equation}
holds for any distribution $P(N)$, then we will have a complete de-biasing prescription for the codec.

When the distribution $P(N)$ is wider than a code step width $\Delta$, it can be approximated by the Taylor expansion of $P(N)$ about $N_i$:
\begin{equation}
P(N_i+u) = P_i + \Pp_i u + \Ppp_i u^2/2 + \Pppp_i u^3 / 6 + O(u/\sigma_N)^4,
\end{equation}
with $\Pp_i = \Pp(N_i)$, etc.  We will truncate the series at cubic order henceforth.  The contribution of the neglected terms will scale as $(\Delta / \sigma_N)^4=b^{-4}$.  We will later perform a numerical check of the significance of the neglected terms. Substituting the Taylor expansion into \eqq{bias1} gives
\begin{equation}
\delta \bar N \approx -\sum_i \Delta_i \left[ \Pp_i \Delta_i^2/12
 + \Pppp_i \Delta_i^4/480 \right].
\label{bias3}
\end{equation}

If this sum over $i$ were an integral, we would evaluate this quantity through integration by parts.  We can perform an analogous discrete process by defining $P_{i\pm1/2}\equiv P(N_i\pm\Delta_i/2)$ and noting that to our chosen order of Taylor expansion, we have
\begin{equation}
\Delta_i \Pp_i  =  P_{i+1/2}-P_{i-1/2}- {\Delta_i^3 \over 24}\Pppp_i.
\end{equation}
Substituting this into \eqq{bias3} yields
\begin{eqnarray}
\delta \bar N & = & -\sum_i \left[ {\Delta_i^2 \over 12} \left(P_{i+1/2}-P_{i-1/2}\right)
  - {\Delta_i^5 \over 720} \Pppp_i \right] \\
 & = & \sum_i \left[ {\Delta_i \over 12} \left( \Deltap_i P_{i-1/2} + \Deltap_{i+1} P_{i+1/2} \right)
  + {\Delta_i^5 \over 720} \Pppp_i \right].
\label{bias4}
\end{eqnarray}
Recalling that $\Deltap_i=\Delta_i-\Delta_{i-1}$, it is clear first that {\em the codec bias is very small if the step width $\Delta$ is constant ($\Deltap=0$).}  This would be the case if the codec were a simple decimation over the range of the data.  This is the case for normal digitization of the data.

The perfect square-root compressor of real-valued raw data  \eqq{encode} has the unique property that $\Deltap_i=\Deltap$ is constant.  In this special case we can simplify further and used the Taylor expansion one more time to get $P_{i+1/2}+P_{i-1/2}=2P_i+\Delta_i^2\Ppp_i/4,$ and now we have
\begin{equation}
\delta \bar N = \sum_i \Delta_i \left[
{\Deltap \over 6} \left( P_i + {\Delta_i^2 \over 8} \Ppp_i\right) 
  + {\Delta_i^4 \over 720} \Pppp_i \right].
\end{equation}

Second, it is clear that {\em the square-root codec bias will be nearly independent of $P(N)$} because $\sum_i \Delta_i P_i\rightarrow 1$ in the limit of small $\Delta_i$, so we expect there to be bias
\begin{equation}
\delta \bar N = {\Deltap \over 6} = {1 \over 12 b^2 g}.
\label{realbias}
\end{equation}
A constant bias is relatively benign, for example does not influence the appearance of a sky-subtracted image.  It is also easily remedied by decompressing code $i$ to value $N_i-\delta^{(0)}$, where $\delta^{(0)}=\Deltap/6$.  After making this adjustment, the residual bias, to third order in the Taylor expansion of each code range, becomes
\begin{equation}
\delta \bar N = \sum_i \Delta_i \left[ {\Deltap \Delta_i^2 \over 72} \Ppp_i
  + {\Delta_i^4 \over 720} \Pppp_i \right].
\end{equation}
Further application of the ``integration by parts'' technique shows that the leading terms in this residual bias scale as $\int dN\, \Deltap \Pp(N)$, which is zero if $\Deltap$ is constant across $N$.  Hence we expect the reconstruction bias to be very close to the constant $1/12b^2g$ for the reconstruction of real-valued raw data with $b\ge 1$, {\it i.e.} whenever the intrinsic noise spans multiple compression code steps.

To summarize the key results: a simple decimation codec ($\Delta_i=$constant) reproduces the mean of the input data with no bias, and the square-root codec ($\Deltap_i=$ constant) biases the mean by the constant value in \eqq{realbias} independent of the distribution $P(N)$ of the raw data.  These statements are very good approximations when $b\ge1$ and the raw data are real-valued.

\subsection{Integer data}
Next we consider the case when the raw data $N$ must be integral.  In this case $\Delta_i$ and $\Deltap_i$ must be integral, and $N_i$ is integral or half-integral depending upon whether $\Delta_i$ is odd or even, respectively.  The integrations over $N$ in \eqq{bias1} must be replaced by sums over the allowed integral values of $N$.  We obtain
\begin{equation}
\delta \bar N = \sum_i \sum_j
j\, P_{N_i+j},
\label{nbias1}
\end{equation}
where the sum runs over $j \in \{-(\Delta_i-1)/2, -(\Delta_i-3)/2, \ldots, (\Delta_i-3)/2, (\Delta_i-1)/2\}$.

We smoothly extend the discrete probabilities $P_N$ to a continuous function $P(N)$ that maintains the normalization condition $\int dN\, P(N) = 1$, and we again perform a 3rd-order Taylor expansion of $P(N)$ about the center $N_i$ of each code range.  The bias can now be written
\begin{equation}
\delta \bar N = -\sum_i \left[ \Pp_i \sum_j j^2 + {\Pppp_i \over 6} \sum_j j^4 \right].
\end{equation}
The sums of powers of integers (or half-integers) can be found from well-known formulae.\footnote{For example, {\tt http://mathworld.wolfram.com/FaulhabersFormula.html}} The result is
\begin{equation}
\delta \bar N = -\sum_i {\Delta_i \over 12} \left(\Delta_i^2-1\right)
\left( \Pp_i + {3\Delta_i^2 - 7 \over 120} \Pppp_i\right).
\end{equation}
The discrete version of integration by parts transforms this into an expression identical to \eqq{bias4} up to small corrections in the $\Pppp$ term:
\begin{equation}
\delta \bar N = \sum_i {\Delta_i \over 12} \left[
 \Deltap_i P_{i-1/2} + \Deltap_{i+1} P_{i+1/2} 
  + {(\Delta_i^2-1)(2\Delta_i^2+7) \over 120} \Pppp_i \right].
\label{nbias2}
\end{equation}
For an integer-valued codec we cannot necessarily maintain constant $\Deltap_i$ for an arbitrary $b$; for example with $g=1$ and $b=1$, we must have $\langle \Deltap \rangle=1/2$.  Hence we must choose adjustments $\delta_i$ to each decompressed value that will best cancel the bias in \eqq{nbias2}.  After further manipulation it is possible to show that the leading-order terms in the bias are cancelled, independent of the form of $P_N$, if
\begin{eqnarray}
\delta_i & = & { \Deltap_{i+1}+\Deltap_i \over 12}
 - { \Deltap_{i+2}-\Deltap_{i+1} - \Deltap_i + \Deltap_{i-1} \over 48} \\
 & = & { \Delta_{i+1}-\Delta_{i-1} \over 12}
 - { \Delta_{i+2}-2\Delta_{i+1}  + 2\Delta_{i-1} -\Delta_{i-2} \over 48}.
\label{nbias3}
\end{eqnarray}
To summarize:  the compression algorithm is defined by a repeating integer sequence $\{s_i\}$ for the $\Deltap$ values, along with a gain $g$ and read noise $RN$, as described in \S\ref{stepseq}.  The compressed values will have $b=(2g \langle s_i \rangle)^{-1/2}$ bits per $\sigma$ of input noise.  On decompression, each compressed code $i$ should be replaced with the value $N_i-\delta_i$, with $\delta_i$ given by \eqq{nbias3}.  

One should prefer the most uniform possible sequences $\{s_i\}$.  A constant value (nominally $\Deltap=1$) will yield a ``perfect'' square-root compression with minimal reconstruction bias.  An alternating sequence, {\it i.e.} $\{0,1\}$, has a constant bias correction $\delta_i$, but slightly larger biases.  Longer $\{s_i\}$ sequences have more complicated and less precise bias corrections.

\subsection{Numerical validation}
\begin{figure}
\plotone{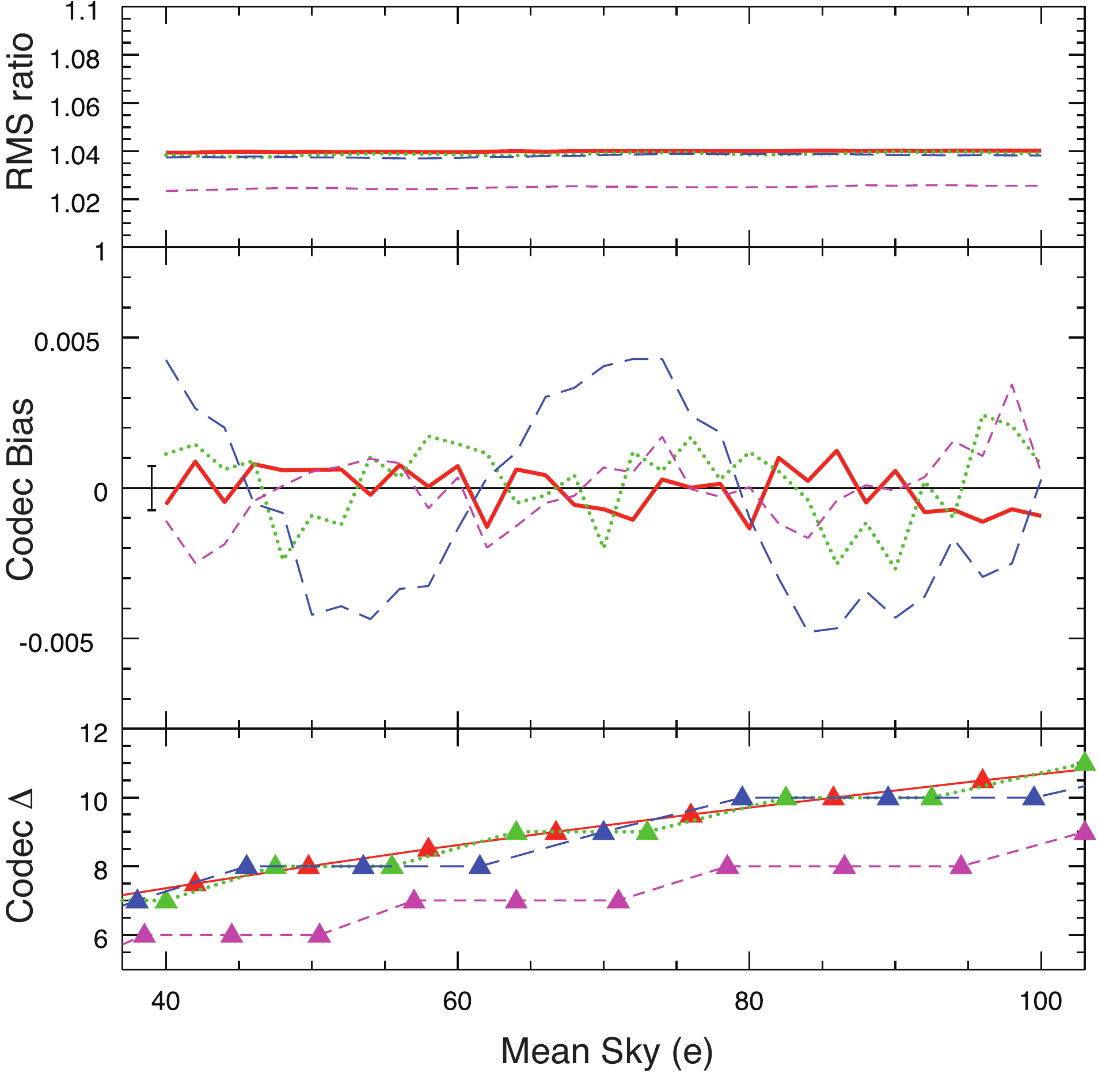}
\caption[]{
\small
Performance of four square-root codecs on data with Poisson noise plus $4e$ RMS Gaussian read noise.  The bottom panel shows the center values $N_i$ and the range widths $\Delta_i$ of each step for the four codes.  The red (solid), green (dotted), and blue (long dash) codecs each produce $b=1$ bit per noise $\sigma_N$ of the input value, with decreasing levels of smoothness in the derivative $\Deltap$ of the range width.  The magenta (short dash) codec retains more noise, $b=1.22$.  The top panel shows the increase in data RMS due to the codec, which is independent of input level.  The central panel shows the bias in the reconstructed data stream, with the noise level of this measure indicated by the black error bar.  Bias levels are $O(0.001e)$, in some cases consistent with zero within measurement noise.  See text for further detail.
}
\label{compress3}
\end{figure}

Figure~\ref{compress3} shows the noise and bias introduced by several versions of the square-root codec.  
In each case the raw data are drawn from a Poisson distribution with the mean number of photoelectrons plotted on the $x$ axis, ranging from 40--100$e$.  The data are taken to have Gaussian read noise with zero mean and $4e$ RMS.  Each plotted point is the result of $10^7$ trials from this distribution.  Each trial produces an integral raw $N$ value and a real-valued decompressed value $N_i-\delta_i$.

We examine 4 different codecs.  The red, blue, and green curves are designed to produce $b=1$ bit of noise.  The red curve is a ``perfect'' square-root codec, with $\Deltap=1$ and $g=0.5e$ per ADU.  The blue and green curves have $g=1$ and $\{s_i\}$ sequences of $(0,1)$ and $(0,0,1,1)$, respectively.  The magenta curve is a codec with $g=1$ and step sequence $(0,0,1)$, which retains $b=1.22$ bits of noise.  

The bottom panel plots the mean $gN_i$ and range $g\Delta_i$ of each code step $i$ for each of the four codecs, expressed in units of photoelectrons.  The top panel shows the RMS fluctuation of the decoded data, relative to the RMS fluctuation of the input data.  All of the codecs inflate the RMS by almost exactly the factor $(1+1/12b^2)^{1/2}$ predicted in \eqq{noiseeq}.  In particular the $b=1$ codecs inflate the noise by 4\%.

The central panel plots the mean difference between decompressed and raw data.  The expected noise in this measurement from the $10^7$ trials is plotted in black at left.  The ``perfect'' codec, in red, has zero bias within errors, $|\delta \bar N| < 0.001$.  This bias is, remarkably, $<10^{-5}$ of the input values, $<10^{-4}$ of the typical noise $\sigma_N$ and code range $\Delta=$7--10, and also $<1\%$ of the bias $1/12b^2=0.083$ that the codec would have generated without any correction.  The $\{0,1\}$-sequence, in green, shows some significant residual bias at the $\pm0.002$ level, and the $\{0,0,1,1\}$-sequence codec (blue) has bias of $\approx\pm0.004$ electrons.  While these biases are quite small, they do highlight that \eqq{nbias3} ignores some higher-order terms that make the bias dependent on $P_N$, and that it is preferable to use the smoothest possible step sequences.  The $\{0,0,1\}$ codec also has some bias at the 0.001 level, despite having slightly less compression and lower noise than the ``perfect'' $b=1$ codec in red.

These numerical tests insure that square-root compression at $b=1$ will introduce no significant biases into photometric measurements, and increases in errors will be limited to 4\%.

\section{Shape measurement biases}
\label{testsec}
Using simple simulated galaxies and a very basic shape-measurement routine, we demonstrate here that the codec does not induce any biases in the measurements of galaxy fluxes, sizes, or ellipticities at the part-in-$10^4$ level.  In this work we analyze simple galaxies with specified size and signal-to-noise ratios which span a range that we might expect in real images.  A test for codec with much more realistic images is presented by \citet{Ali}.  In that work the galaxies have sizes, appearances, fluxes, and shapes that accurately mimic those we might expect from a visible survey telescope in orbit.

\subsection{Procedure}
The simulated images are postage stamps constructed as follows:
\begin{enumerate}
\item Start with an exponential-profile galaxy, with intensity $I\propto e^{-r/r_0}$.  The exponential disk is convolved with a Gaussian of width $\sigma=r_0/10$, to suppress high frequencies associated with the $r=0$ cusp of the exponential disk.  
\item This circular galaxy is then sheared to an ellipticity $e=(a^2-b^2)/(a^2+b^2)=0.1$, 0.3, or $0.7$ ($a$ and $b$ are the major and minor axes of the galaxy).  The major axis of the elliptical galaxy can have position angle $\phi=0$, 22.5\arcdeg, or 45\arcdeg\ from the $x$ axis.
\item The galaxy is (de-)magnified until the resulting galaxy has half-light radius of $r_{1/2}=2$, 4, or 6 pixels.
\item The galaxy profile is shifted by a random fraction of a pixel in $x$ and $y$, then convolved with the square pixel response function and sampled on a 1-pixel grid to yield a noiseless postage stamp image.
\item The flux of the galaxy is adjusted until its total signal-to-noise ratio will be $S/N=15$, 50, or 300 when noise is added in the following step.  The $S/N$ is defined as
\begin{equation}
(S/N)^2 = \sum_{i \in {\rm pixels}} { I^2_i \over \sigma^2_i}
\end{equation}
where $I_i$ and $\sigma_i$ are the signal and noise in each image pixel.
\item A sky level of 50 electrons is added to each pixel's expected signal $I_i$.  Then an integral ``observed'' photoelectron count for each pixel is determined by drawing from a Poisson distribution with mean $I_i$.  Gaussian read noise with zero mean and $\sigma=4$ electrons is added to give a floating-point measurement value for each pixel.
\end{enumerate}

At this point two digitized versions of the postage-stamp image are produced.  For the ``codec'' image, the photoelectron count $I_i$ is scaled by gain $g=0.5e/$ADU, digitized, compressed with a sequence ${1}$ perfect square-root codec, and then decompressed to a floating point image following the debiasing procedures above.  Note that the codec retains $b=1$ bits per noise sigma on each pixel.  For the ``raw'' image, we add Gaussian noise with RMS of $\sigma_i/\sqrt{12}b$ to each pixel, so that the noise level of the raw and codec images will be expected to be equal.  Then we scale by the gain $g$ and digitize {\em without} square-root preprocessing.

Both the raw and codec versions of the postage stamp galaxy are measured using a basic ``adaptive moments'' shape-measurement algorithm \citep{BJ02} as follows: the pixel data are fit to an elliptical Gaussian model, with the flux, centroid, size, and ellipticity of the Gaussian allowed to vary.  The sky background is fixed at $50/g=100$ ADU.  There are hence 6 degrees of freedom in galaxy model (and the Gaussian galaxy model is {\em not} an exact fit to the underlying exponential-profile galaxy).  

\subsection{Results}
For each choice of ellipticity $e$, orientation $\phi$, size $r_{1/2}$ of galaxy, we create 1000 simulated galaxies at $S/N=300$, 30,000 realizations with $S/N=50$, and 300,000 realizations with $S/N=15$.  To test for biases induced by the codec,  we measure the mean difference $e_{\rm codec}-e_{\rm raw}$ between the ellipticities measured on the codec and the raw images.  Similarly we measure the mean difference in the (log of) flux and size measured for the galaxies.  To see if the codec induces extra shape noise, we compare the RMS variation of the fitted $e$ in the codec data set to the RMS variation in the raw fits.  We also check for additional variance in flux and size measurements.

\begin{figure}
\plotone{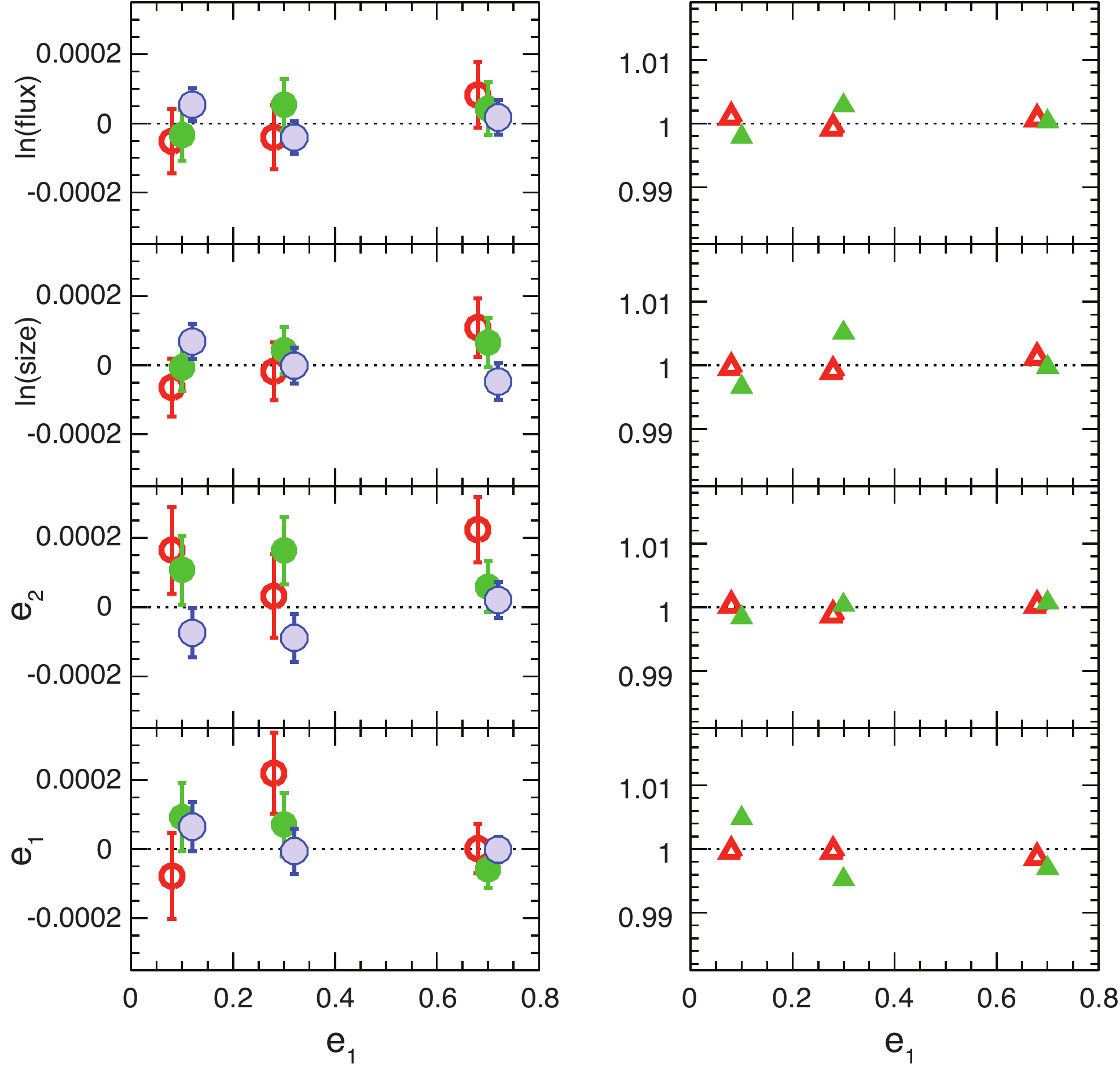}
\caption[]{
The biases (left) and relative noise level (right) of measurements on codec images, relative to raw images, are plotted for several quantities measured on simulated galaxies.  The details of the simulation and the measurement process are in the text.  From top to bottom, we plot the bias in the (log of) galaxy flux, (log of) galaxy size, and then the two components of galaxy ellipticity.  Red (outlined) points are for galaxies with $S/N=15$, green (filled) have $S/N=50$, and blue (shaded) have $S/N=300$, with each point plotted at its input ellipticity.  All data are fully consistent with the codec images yielding no bias or change in noise level from the raw images---at accuracies of $\pm10^{-4}$ for biases and $<1\%$ in noise levels.  [Noise levels for $S/N=300$ cases are not plotted because there are too few trials for an accurate measure of RMS noise.]
}
\label{emeans}
\end{figure}

Figure~\ref{emeans} presents the biases and noise levels on the ellipticity, size, and flux measurements of galaxies with $r_{1/2}=2$ pixels and $\phi=0$.  The results for larger galaxies and for different ellipticity orientations are similar and not plotted here.  We find no evidence for any codec-induced bias in size, flux, or ellipticity measurements at any $S/N$ or ellipticity.  The number of trials is sufficient to determine these biases to $\approx \pm10^{-4}$ for each configuration, approximately 10 times below the maximum permissible ellipticity measurement errors in a large WL survey \citep{AmaraRefregier, HTBJ}.  For these simplified galaxies and shape measurement techniques, we confirm that the square-root codec at $b=1$ does not induce any significant bias in weak-lensing shape measurements.

The $S/N=50$ and $S/N=15$ cases have enough trials to determine the measurement noise in $e$, size, and flux to $\ll 1\%$.  We find the raw and codec images to have identical measurement noise levels to this accuracy, confirming that the noise added by the codec is accurately described by \eqq{noiseeq}.

\section{Compression factors}
\label{bppsec}
\subsection{Simulated scene}
We create simulated images from a space-based visible survey telescope to estimate the bits of telemetry required per pixel (bpp) needed to transmit the images after they have been compressed with the square-root algorithm followed by a lossless compression algorithm.  The simulated images have three components: the expected cosmic scene; a uniform sky background; and bright linear features produced by cosmic rays in the detectors.  Our test images are aimed at reproducing the appearance of images taken with an instrument resembling the CCD camera proposed for the {\it JDEM/IDECS} mission. The relevant characteristics assumed are: a 1.5-meter diameter telescope operating in an L2 Lissajous orbit, where the sky background is dominated by zodiacal light; a 200-second integration through a bandpass in the red part of the visible spectrum, with bandwidth $\ln(\lambda_{\rm max}/\lambda_{\rm min})=0.3$; a fully-depleted, 200-micron-thick p-channel CCD \citep{Roe}; a scale of 0\farcs14 per pixel; and a point spread function (PSF) with EE50 radius $\approx0\farcs14$, after including all optical, mechanical, and detector-induced sources of image blur.

The cosmic scene is taken from deep moderate-latitude exposures taken with the ACS/WFC F606W filter aboard the Hubble Space Telescope in the course of a search for very faint outer-solar-system objects \citep{acskbo}.  These exposures combine many HST exposures to produce an image whose noise is far below that expected from a single {\it JDEM/IDECS} exposure as taken above.  The scene is scaled so that a source of magnitude $I_{AB}=31.09$ produces 1 photoelectron.  The HST image is blurred and repixelized to match the {\it JDEM/IDECS} specifications described above.

Diffuse sky emission is added assuming high-ecliptic-latitude zodiacal background at L2.  We estimate $\approx 45$ electrons per pixel in the detector.

\subsection{Simulated cosmic rays}
Cosmic rays in space are isotropic and uniformly impinge on a sphere.  From ESA's Space Environment Information System (SPENVIS)\footnote{\tt http://www.spenvis.oma.be/spenvis/}  program, the cosmic ray rate is found to be approximately 5 particles~s$^{-1}$~cm$^2$ over $4\pi$ steradian.  This is the average rate for Galactic cosmic rays, which are dominated by protons.  Low energy particles (mainly solar protons) have little integrated effect on the total rate even without shielding.  We ignore periods of enhanced low momentum solar protons during solar storms.  These episodes are known in advance and data taking is avoided.

We use SPENVIS to generate a momentum spectrum for a particular orbit (L2 in this case) at a particular solar epoch behind appropriate shielding (3 mm Al assumed here).  The Monte Carlo simulation samples this spectrum and for each momentum, an average ionization rate (e/$\mu$m) is calculated for the target material (Si in this case) according to a modified Bethe-Bloch formalism \citep[Section 27]{PDG}.  For the data generated and analyzed here, a single particle momentum was used corresponding to minimizing protons---approximately 3 GeV/c corresponding to 70 e/$\mu$m in silicon.

For a given exposure time, particle flux and detector area, the total number of cosmic rays is calculated.  We generate this number of cosmic rays uniformly over $4\pi$ sr.  Each ray is randomly placed over the surface of detector and tracked through the detector material.  Tracing is done in uniform step lengths.  For each step, a Landau distribution is sampled to determine the number of ions produced.  Hard scatters and $\delta$-rays are not simulated.  For each ion generated in a step, a Gaussian lateral charge diffusion distribution is randomly sampled.  The diffusion constant is based on a measured 4 $\mu$m RMS for a 200 $\mu$m thick CCD operated at 100V bias voltage.  This is linearly scaled by ion conversion height above the pixel plane.  The diffused charge is projected to the pixel plane where its position is discretized and added to any previous pixel charge.

In the simulated 200~s exposure, 7\% of the pixels receive $\ge5e$ of CR-generated charge.  We also simulate an image with triple this dose of cosmic rays, which is simply the sum of 3 200s CR simulations.

\subsection{Compression results}
\begin{figure}
\plotone{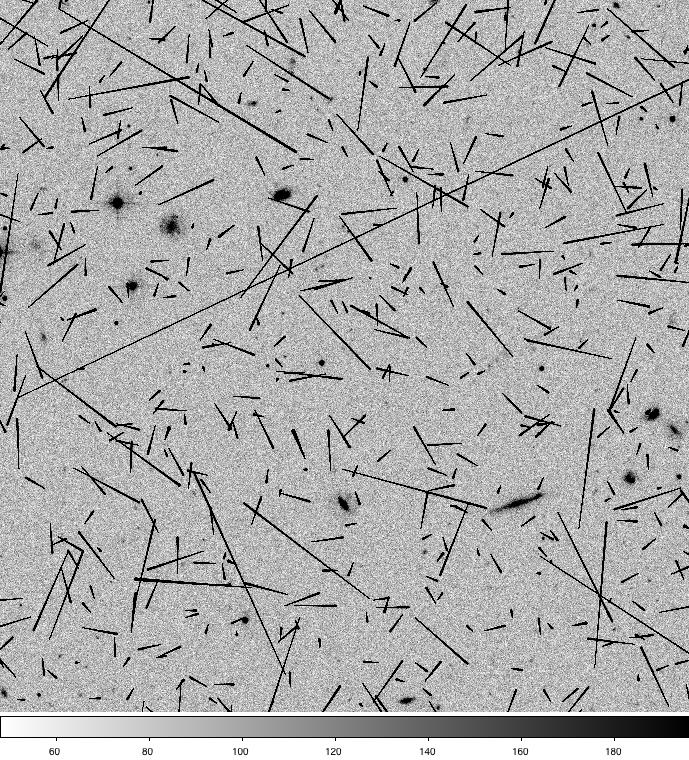}
\caption[]{Section of an image used to test the square-root$+$lossless compression scheme.  This simulates a 200-second CCD exposure with a 1.5-meter space telescope observing near the ecliptic pole from an L2 orbit.  Most pixels are dominated by sky and read noise, and celestial objects are visible, but the cosmic-ray tracks dominate the information content of the image.  The scale is in ADU, and the image gain is 0.5 photoelectrons per ADU.  The image requires 3.1 bits per pixel to transmit after application of square-root compression and lossless {\tt bzip2} compression.}
\label{image}
\end{figure}

The scene, cosmic ray, and sky components are summed, and then Poisson noise and $4e$ RMS read noise are added to each pixel.  The noisy images are then scaled by gain $g=0.5e/$ADU and digitized, then compressed using a perfect square-root algorithm with sequence ${1}$.  The compressed images are saved as 16-bit integer FITS-format images.  A segment of a simulated image, before compression, is shown in Figure~\ref{image}.

The Shannon entropy of the square-root-compressed image, calculated by direct measure of the $p_i$ of each code in the image, is 2.8 bits per pixel.  Tripling the cosmic-ray dose on the exposure increases the entropy to 3.6 bits per pixel, suggesting that the normal dose of cosmic rays is responsible for 0.4 bits per pixel of entropy in the image.  
Thus most of the 0.7-bit increase in entropy over the pure-noise image (as measured in \S\ref{noisesec}) is due to cosmic rays.  

There are many possible lossless compression algorithms to apply to the square-root-compressed FITS images.  This choice will of course not affect the biases in the data, but will affect the compression ratio and the robustness of the telemetry stream.  Applying the commonly available {\tt bzip2} algorithm to the FITS files yields the following results:
\begin{itemize}
\item The images simulating a 200-second exposure and 200 seconds' worth of cosmic-ray tracks occupy {\bf 3.1 bits per pixel} after square-root and {\tt bzip2} compression.
\item The information content of the images is dominated by the cosmic-ray component.  If we triple the number of cosmic rays (mimicking a 600-second exposure, even though the cosmic scene and sky background are still 200 seconds' worth), we get a pessimistic value of {\bf 4.0 bits per pixel} in the compressed images.
\end{itemize}
The {\tt bzip2} performance is again 0.3--0.4 bpp above the entropy of the image.

Full-frame {\tt bzip2} compression is not very robust: a single-bit transmission error can lead to loss of a full image.   A more robust algorithm is CCSDS 121B \citep{ccsds}, which has already been implemented on $>25$ space missions.  We find the file size after square-root pre-processing and CCSDS 121B lossless compression to be within 0.1 bits per pixel of the {\tt bzip2} results.

A different lossless compression scheme, optimized toward efficient compression of the linear cosmic-ray tracks that dominate the 2d image, might yield better compression yet.
If further compression were desired, the most effective approach would be to identify cosmic-ray-contaminated pixels on that satellite and replace them with a constant tag value before compression, so that their entropy is not transmitted to the ground. 

\subsection{Comparison to linear preprocessing}
Implementation of square-root preprocessing is very simple, requiring just a fixed lookup table to convert ADC output values to compressed codes.  An even simpler algorithm is  linear prescaling, {\it i.e.} decimation, in which the code range $\Delta$ is fixed over the full dynamic range.  As noted above, a decimation algorithm should be free of bias, just as the corrected square-root algorithm is measured to be free of bias.  What are the relative merits of the two lossy codecs?

We compress our test images with a linear algorithm with code step $\Delta$ chosen to yield $b=1$ steps per noise $\sigma$ at the sky level of the images.  As noted {\it e.g.} in \citet{PSW}, most pixels are at sky level so a fixed $b$ at sky level should yield similar compression performance.  Indeed for images with {\em no} cosmic rays, we find that linear and square-root compression yield indistinguishable bpp values after the lossless step.  As expected, the square-root algorithm performs better if the number of CR-impacted, high-signal pixels is substantial.  In our nominal 200s exposure, the square-root algorithm uses 0.12 fewer bpp than decimation.  For a 600s CR dose, the square-root algorithm requires 0.38 fewer bpp than decimation.  This suggests that the square-root algorithm removes about 2 bits of noise from the average CR-impacted pixel.  If an algorithm to remove cosmic rays before compression is implemented, then there would be little difference in compression ratio between the square-root and linear codecs.  However such an algorithm would require more processing power than the lookup table of the square-root codec, and one would have to insure that objects of interest are not discarded with the cosmic rays.

 We should expect the square-root algorithm to gain advantage also in the case of fields with many objects brighter than the sky background, {\it e.g.} fields at low Galactic latitude or with large nearby galaxies.

The square-root algorithm has practical advantages as well, in that it will retain the desired $b=1$ value as the sky background varies.  The linear codec would need to change its step size $\Delta$ to accommodate different sky levels, {\it e.g.} from changing filters or zodiacal background.  An on-board algorithm could determine the sky noise level and choose $\Delta$ appropriately for the linear codec.  The square-root codec requires adjustment only when the bias level or gain of the detector ADC output change.

\section{Conclusion}
Future dark-energy surveys such as {\em JDEM} or {\em Euclid}\footnote{\tt http://sci.esa.int/euclid} propose to 
survey a substantial fraction of the full sky with space-based visible/NIR imaging near the diffraction limit.  A data compression scheme is needed to satisfy telemetry constraints, but the weak gravitational lensing measurements place very strict requirements on the degree to which the compression algorithm can bias the measured shapes or fluxes of galaxies.  For this reason we choose to investigate lossy compression processes that operate on a pixel-by-pixel basis, fearing than algorithms examining pixel groups might induce correlated codec errors between pixels which bias the galaxy ellipticities.

Square-root pre-processing is known to be a good means for compressing shot-noise-dominated data streams while keeping a fixed number $b$ of bits per $\sigma$ of noise across the full detector dynamic range.  The reduction of entropy by reducing the number of noise bits is essential to efficient subsequent lossless compression algorithms.  We derive an expression for bias in the naive reconstruction of square-root-compressed data and present a correction scheme, which we numerically verify is successful to high accuracy.

Since the noise induced by roundoff in the compression process increases the required observing time by $1/12b^2$, we confine our examination to $b\ge1$.  We verify that flux and shape measurements of simple galaxy models are biased by $\lesssim 10^{-4}$ for a range of galaxy sizes, ellipticities, orientations, and signal-to-noise ratios.  This is well within the weak lensing requirements.  We verify this with simple postage-stamp galaxy images; \citet{Ali} bound the shape biases determined from realistic simulations of the true extragalactic sky.

At $b=1$ the Shannon entropy of the square-root-compressed images varies from 2.1 bits per pixel for pure noise to 3.6 bits per pixel for images with the expected astronomical scene and triple the expected number of cosmic rays.  Lossless compression algorithms such as {\tt bzip2} realize compress levels within 0.3--0.4 bits of the Shannon limit.   For simulated images best resembling expected {\em JDEM} CCD data, we find compression to 3.1 bits per pixel using {\tt bzip2}.  The size of the transmitted data stream depends primarily on the degree of cosmic-ray contamination once $b=1$ is set: schemes to more efficiently compress or mask cosmic rays before compression might lead to lower bit rates.  Higher cosmic-ray rates or more crowded astronomical fields, {\it e.g.} near the Galactic plane, will require more bits.  Since pure noise images require 2.4 bits per pixel with {\tt bzip2} at $b=1$, we can state that compression to $<2.4$ bits per pixel will require $b<1$, and hence higher observing-time penalty.

We thus validate square-root lossy compression as useful for {\em JDEM} data and capable of meeting requirements for low bias.  Other approaches would be feasible as well. For instance simple decimation to a fixed code step $\Delta$ that equals the sky noise level should produce similar noise and bias properties.  The presence of cosmic rays and/or bright objects in $\gtrsim 10\%$ of the pixels gives the square-root codec a useful advantage in mean number of required telemetry bits per pixel.  The square-root algorithm also has a robust and exceptionally simple implementation---a single lookup table---meaning that there is no need for on-board processing to recognize cosmic rays or determine the sky noise level.

It is thus expected that visible/NIR dark-energy extragalactic imaging survey data can be transmitted to the ground with 2.5--4 bits per pixel, depending upon the cosmic-ray flux level and the sophistication of onboard cosmic-ray identification or removal.  This compression can be achieved with no significant bias of observed quantities, and with 8\% penalty in observing time to achieve fixed signal-to-noise ratio.  

\acknowledgements
The work of JR and RAV was carried out at the Jet Propulsion
Laboratory, California Institute of Technology, under a contract with
NASA, and was funded by JPL's Research and Technology Development
Funds.  GB is supported by grant AST-0607667 from the NSF and DOE grant
DE-FG02-95ER40893.  CB is supported by the Director, Office of Science, Office of High Energy Physics, of the U.S. Department of Energy under Contract
No. DE-AC02-05CH11231. Fermilab is operated by Fermi Research Alliance, LLC under Contract
No. DE-AC02-07CH11359 with the United States Department of Energy.

\end{document}